\documentclass[10pt,fleqn]{article}
\usepackage[T1]{fontenc}
\usepackage[utf8]{inputenc}
\usepackage[a4paper]{geometry}
\usepackage{color}
\usepackage{bm}
\usepackage{amsmath}
\usepackage{amssymb}
\usepackage{setspace}
\usepackage{esint}
\usepackage[unicode=true,pdfusetitle,
 bookmarks=true,bookmarksnumbered=false,bookmarksopen=false,
 breaklinks=false,pdfborder={0 0 1},backref=false,colorlinks=true,pdfpagemode=FullScreen]
 {hyperref}
\usepackage{breakurl}


\begin{document}
$ $

\begin{center}
\textbf{\LARGE \vspace{1mm}
}
\par\end{center}{\LARGE \par}

\begin{center}
\textbf{\LARGE Involutive constrained systems and Hamilton-Jacobi
formalism}
\par\end{center}{\LARGE \par}

\begin{center}
\textbf{\LARGE \vspace{1mm}
}
\par\end{center}{\LARGE \par}

\begin{center}
M. C. Bertin%
\footnote{mbertin@ufba.br%
}, B. M. Pimentel%
\footnote{pimentel@ift.unesp.br%
}, and C. E. Valcárcel%
\footnote{carlos.valcarcel@ufabc.edu.br%
}, 
\par\end{center}

\begin{center}
$^{1}$\emph{Instituto de Física, Universidade Federal da Bahia},\\
 \emph{Câmpus Universitário de Ondina, CEP: 40210-340, Salvador -
BA, Brazil}.
\par\end{center}

\begin{center}
$^{2}$\emph{Instituto de Física Teórica, UNESP - São Paulo State
University},\\
 \emph{Caixa Postal 70532-2, 01156-970, São Paulo, SP, Brazil}.
\par\end{center}

\begin{center}
$^{3}$\emph{CMCC, Universidade Federal do ABC},\\
 \emph{Rua Santa Adélia, 166, Santo André, SP Brazil}. 
\par\end{center}
\begin{abstract}
\thispagestyle{empty}In this paper, we study singular systems with
complete sets of involutive constraints. The aim is to establish,
within the Hamilton-Jacobi theory, the relationship between the Frobenius'
theorem, the infinitesimal canonical transformations generated by
constraints in involution with the Poisson brackets, and the lagrangian
point (gauge) transformations of physical systems.
\end{abstract}
\tableofcontents{}

\section{Introduction\label{sec:1}}

$ $

Symmetries become the cornerstone of the modern physics, from the
advent of the special theory of relativity to the standard model of
the elementary particles. Among the physical requirements for the
construction of fundamental interactions in quantum field theory is
the concept of gauge symmetry and, for the gravitational field, the
concept of diffeomorphism invariance in a Riemannian space-time. In
the mathematical sense, integrability of ordinary (ODEs) and partial
differential equations (PDEs) is usually related to the existence
of certain integrals of motion of dynamical systems, and when the
physical systems to which those equations belong can be described
by variational problems, symmetries are found to be related to the
existence of these integrals of motion. The astounding beauty of the
subject is that mathematical symmetries determine the physical interaction
of the matter fields in nature.

The study of symmetries in field theory is historically linked to
the lagrangian and hamiltonian formalisms. In the lagrangian picture
we have the Noether's theorems, which relate symmetries of a fundamental
integral to conserved currents and geometrical identities in the context
of the calculus of variations. The hamiltonian formalism, on the other
hand, deals with symmetries in the context of canonical transformations.
Even in classical mechanics, where both formalisms are completely
equivalent, there is no general and unique correspondence between
canonical and lagrangian symmetries.

In field theories, the existence of lagrangian symmetries implies
singularity of the lagrangian function, which means that there are
constraints relating some of the phase-space variables \cite{Bergmann,Anderson}.
In this case, the hamiltonian picture must be constructed to be equivalent
to the lagrangian one. The first and most used method to build this
equivalence is Dirac's hamiltonian method \cite{Dirac}, which consists
in the construction of the hamiltonian picture by consistency. For
detailed textbooks on the subject, we refer to \cite{Sunderm}.

Dirac found that symmetries of a singular system are related to the
existence of a subset of the so called primary first-class constraints,
which are generators of \textquotedbl{}gauge transformations\textquotedbl{}
in the hamiltonian method. However, this relationship has no general
rule in sight of the applications, and he found himself obligated
to conjecture that the complete generator of the lagrangian symmetries
of a system must be a linear combination of all first-class constraints
\cite{Dirac1}. In the mathematical point of view, this problem is
still considered unsolved, despite some results in the literature
\cite{Dirac conj}, and the success in applying the conjecture in
the analysis of gauge theories \cite{Castellani}.

On the other hand, the Hamilton-Jacobi (HJ) formalism provides a very
natural way of dealing with symmetries, and also with singular systems.
Carathéodory \cite{Cara} was the first to notice that this formalism
is the unifying theory between the theory of first-order PDEs, the
theory of first-order ODEs, and the calculus of variations. Lagrangian
and hamiltonian dynamics find place as direct mathematical consequences
of the HJ theory. With the suggestive name of \textquotedbl{}the complete
figure\textquotedbl{} of the variational calculus, the HJ formalism
was extended to treat singular systems by Güler \cite{Guler}, followed
by generalizations for higher order derivative lagrangians \cite{gen},
Berezin systems \cite{Berezian}, linear actions \cite{FOA} and applications,
specially in the gravitational field \cite{Gravity} and topologically
massive theories \cite{TMYM}. In the HJ formalism, canonical constraints
form a set of PDEs of the first-order, and the dynamical evolution
is generated by a complete set of independent hamiltonian functions,
resulting in a system with several independent variables. In general,
a system presents two different sets of constraints, which are called
non-involutive and involutive constraints. In \cite{GB}, systems
with non-involutive constraints are studied, and it is shown that
these constraints are responsible to change the symplectic structure
of the phase space.

In this paper we provide a continuation of the work \cite{GB}, and
study systems with involutive constraints in sight of the Frobenius'
integrability theorem. In sec. \ref{sec:2} we make a review of the
HJ formalism. In sec. \ref{sec:3} we present an analytic derivation
of the Frobenius' integrability conditions. In sec. \ref{sec:4} we
turn to a geometrical description of the HJ formalism to show that
involutive constraints are generators of canonical transformations
on a complete phase space, also discussing the relationship between
these transformations and the lagrangian transformations in singular
theories. In sec. \ref{sec:5} we present three applications: the
first two are mechanical models for gauge theories, the first one
having just involutive constraints, the second one presenting involutive
and non-involutive ones. The last example is the free Yang-Mills field.

\section{The Hamilton-Jacobi formalism\label{sec:2}}

$ $

In this section we make a brief review of the HJ formalism for singular
systems. Let us consider a system described by the action 
\begin{equation}
I=\int_{t_{0}}^{t_{1}}dt\, L\left(t,q^{i},\dot{q}^{i}\right),\label{eq:01}
\end{equation}
where $q^{i}$ are $N$ generalized coordinates of a configuration
manifold $\mathbb{Q}_{N}$, and $\dot{q}^{i}$ are their respective
velocities. Solutions of a variational problem involving \eqref{eq:01}
are trajectories $q^{i}=q^{i}\left(t\right)$ parametrized by the
time $t$. Although we are dealing with a classical mechanical approach,
extension to field theories is straightforward.

According to Carathéodory \cite{Cara}, the necessary condition for
the existence of an extreme configuration of \eqref{eq:01} is the
existence of a function $S\left(t,q^{i}\right)$ that obeys\begin{subequations}\label{eq:02}
\begin{gather}
\frac{\partial L}{\partial\dot{q}^{i}}=\frac{\partial S}{\partial q^{i}},\label{eq:02a}\\
\frac{\partial S}{\partial t}+\frac{\partial S}{\partial q^{i}}\dot{q}^{i}-L=0.\label{eq:02b}
\end{gather}
\end{subequations}

The HJ formalism emerges by making \eqref{eq:02b} a PDE for $S$.
This can be accomplished if we are able to find expressions of the
velocities $\dot{q}^{i}$ in terms of the coordinates and derivatives
of $S$. Equations \eqref{eq:02a} can be inverted to give such expressions
if the Hessian condition 
\begin{equation}
\det W_{ij}=\det\left(\frac{\partial^{2}L}{\partial\dot{q}^{i}\partial\dot{q}^{j}}\right)\neq0\label{eq:03}
\end{equation}
is satisfied.

However, let us suppose that the Hessian has rank $P\leq N$. This
implies a split of $\mathbb{Q}_{N}$ in two subspaces: the first,
$\mathbb{Q}_{P}$, is spanned by the $P$ coordinates $q^{a}$ related
to the regular part of the Hessian matrix. The second, $\Gamma_{R}$,
is spanned by $R=N-P$ variables $t^{z}\equiv q^{z}$ related to the
null space of $W_{ij}$. Then, we are allowed to invert the equations
for $\dot{q}^{a}$, that give us $P$ velocities 
\begin{equation}
\dot{q}^{a}=\phi^{a}\left(t,t^{z},q^{b},\frac{\partial S}{\partial q^{b}}\right),\,\,\,\,\,\,\,\,\,\,\,\,\,\,\,\,\,\,\, z=1,\ldots,R;\,\,\,\,\, a,b=1,\ldots,P.\label{eq:04}
\end{equation}
The remaining equations cannot be inverted, but they must be valid
nevertheless. We may write them as 
\begin{equation}
\frac{\partial S}{\partial t^{z}}+H_{z}\left(t,t^{z},q^{a},\frac{\partial S}{\partial q^{a}}\right)=0,\,\,\,\,\,\,\,\,\,\,\,\,\,\, H_{z}\equiv-\left.\frac{\partial L}{\partial\dot{t}^{z}}\right|_{\dot{q}=\phi},\label{eq:05}
\end{equation}
We suppose that the functions $H_{z}$ do not depend on $\dot{t}^{z}$,
because otherwise, the relations \eqref{eq:05} would be invertible.
We could allow dependence on $\dot{t}^{z}$ in a non invertible way,
but this dependence would come from very strange lagrangian functions,
which we will not consider here.

Using \eqref{eq:04} in \eqref{eq:02b} we can show that, when \eqref{eq:05}
are obeyed, the hamiltonian function 
\begin{equation}
H_{0}\equiv\frac{\partial S}{\partial t^{z}}\dot{t}^{z}+\frac{\partial S}{\partial q^{a}}\phi^{a}-L\left(t,t^{z},q^{a},\dot{t}^{z},\phi^{a}\right)\label{eq:06}
\end{equation}
does not depend on $\dot{t}^{z}$. Then, \eqref{eq:02b} becomes the
desired PDE 
\begin{equation}
\frac{\partial S}{\partial t}+H_{0}\left(t,t^{z},q^{a},\frac{\partial S}{\partial q^{a}}\right)=0,\label{eq:07}
\end{equation}
known as the Hamilton-Jacobi equation.

Along with \eqref{eq:07}, \eqref{eq:05} are also valid, and together
they form a set of PDEs for $S$. Let us define $t^{0}\equiv t$,
then we are able to write these equations in a unified way: 
\begin{equation}
\frac{\partial S}{\partial t^{\alpha}}+H_{\alpha}\left(t^{\beta},q^{a},\frac{\partial S}{\partial q^{a}}\right)=0,\,\,\,\,\,\,\,\,\,\,\,\,\alpha,\beta=0,1,\cdots,R.\label{eq:08}
\end{equation}
These are the Hamilton-Jacobi partial differential equations (HJPDEs).

\subsection{The canonical description and characteristic equations\label{sub:2.1}}

$ $

In the HJ theory, the conjugate momenta are defined to be in the direction
of the gradient of the function $S$: 
\begin{equation}
\pi_{\alpha}\equiv\partial S/\partial t^{\alpha},\,\,\,\,\,\,\,\,\,\,\,\,\, p_{a}\equiv\partial S/\partial q^{a}.\label{eq:09}
\end{equation}
Now, we define the functions 
\begin{equation}
H'_{\alpha}\left(t^{\beta},q^{a},\pi_{\beta},p_{a}\right)\equiv\pi_{\alpha}+H_{\alpha}\left(t^{\beta},q^{a},p_{a}\right).\label{eq:10}
\end{equation}
In this case, eqs. \eqref{eq:10} are identified with a set of canonical
constraints $H'_{\alpha}=0$. Therefore, the system should be completely
described by the set of HJPDEs\begin{subequations}\label{eq:11}
\begin{gather}
H'_{\alpha}\left(t^{\beta},q^{a},\pi_{\beta},p_{a}\right)=0,\label{eq:11a}\\
\pi_{\alpha}=\frac{\partial S}{\partial t^{\alpha}},\,\,\,\,\,\, p_{a}=\frac{\partial S}{\partial q^{a}}.\label{eq:11b}
\end{gather}
\end{subequations}

The HJ equations \eqref{eq:11a} form a set of $R+1$ PDEs of the
first-order. If $t^{\alpha}$ are independent among each other, we
may find a related set of total differential equations (TDEs)\begin{subequations}\label{eq:12}
\begin{gather}
dq^{a}=\frac{\partial H'_{\alpha}}{\partial p_{a}}dt^{\alpha},\,\,\,\,\,\,\,\,\,\,\, dp_{a}=-\frac{\partial H'_{\alpha}}{\partial q^{a}}dt^{\alpha},\label{eq:12a}\\
dS=p_{a}dq^{a}+\pi_{\alpha}dt^{\alpha}-H'_{\alpha}dt^{\alpha}.\label{eq:12b}
\end{gather}
\end{subequations}These are the characteristic equations (CEs) of
the HJPDEs. Independence between $t^{\alpha}$ is assured by an integrability
theorem, which will be discussed in the next section, but it is important
to remark that full integrability of the HJ equations is a necessary
condition for the derivation of \eqref{eq:12}.

The CEs have the form of canonical equations with several independent
variables $t^{\alpha}$ as evolution parameters. Complete solutions
are congruences of $\left(R+1\right)-$parameter curves 
\begin{equation}
q^{a}=q^{a}\left(t^{\alpha}\right),\,\,\,\,\,\,\,\,\,\,\, p_{a}=p_{a}\left(t^{\alpha}\right)\label{eq:13}
\end{equation}
of a reduced phase space $\mathbf{T}^{*}\mathbb{Q}_{P}$ spanned by
the variables $q^{a}$ and the conjugate momenta $p_{a}$. Observing
\eqref{eq:13}, we call the set $\xi^{A}\equiv\left(q^{a},p_{a}\right)$
the dependent variables of the theory, and the set $t^{\alpha}$ the
independent variables, or parameters. Therefore, it is possible to
describe the dynamical evolution of any function $F\left(t^{\alpha},\pi_{\alpha},q^{a},p_{a}\right)$
in an extended phase space $\mathbf{T}^{*}\mathbb{Q}_{N+1}$ spanned
by the complete set of variables $\xi^{I}\equiv\left(t^{\alpha},\pi_{\alpha},q^{a},p_{a}\right)$.
This is achieved by the fundamental differential 
\begin{equation}
dF=\left\{ F,H'_{\alpha}\right\} dt^{\alpha},\label{eq:14}
\end{equation}
with the extended Poisson brackets (PBs) 
\begin{equation}
\left\{ A,B\right\} \equiv\frac{\partial A}{\partial t^{\alpha}}\frac{\partial B}{\partial\pi_{\alpha}}-\frac{\partial B}{\partial t^{\alpha}}\frac{\partial A}{\partial\pi_{\alpha}}+\frac{\partial A}{\partial q^{a}}\frac{\partial B}{\partial p_{a}}-\frac{\partial B}{\partial q^{a}}\frac{\partial A}{\partial p_{a}}.\label{eq:15}
\end{equation}
The functions $H'_{\alpha}$ are the very generators of the dynamical
evolution \eqref{eq:14}, acting as hamiltonian functions. Therefore,
the HJ formalism describes singular systems as several independent
variables systems.

\section{Integrability\label{sec:3}}

$ $

The HJ equations \eqref{eq:12} become the necessary conditions for
the existence of extreme configurations of the action \eqref{eq:01},
but they are still not sufficient. In deriving the CEs we used the
fact that the independent variables of the system must be mutually
independent. However, this cannot be generally assured only by the
HJ equations. Independence of the parameters is related to the fact
that the evolution of the system in the direction of an independent
variable should be independent of the other variables. On the other
hand, this is related to the very integrability of the theory, which
means the existence of complete solutions of the HJPDEs, as well as
the existence of a unique solution of the CEs once given a set of
initial conditions. In this section, we discuss what are the conditions
that the HJ equations must obey to be a complete integrable system
of PDEs. This can be accomplished in several ways. In our discussion,
we generalize the method presented in \cite{Rund}.

\subsection{The Lagrange brackets\label{sub:3.1}}

$ $

Let us suppose the set of HJ equations \eqref{eq:11a} to be satisfied.
If they form a complete integrable set, there exists a complete solution
with the form 
\begin{equation}
S=S\left[t^{\alpha},q^{a}\left(t^{\alpha}\right)\right].\label{eq:16}
\end{equation}
The function $S$ is submitted to the conditions \eqref{eq:11b},
and the functions $q^{a}=q^{a}\left(t^{\alpha}\right)$, $p_{a}=p_{a}\left(t^{\alpha}\right)$
are supposed to be solutions of the CEs \eqref{eq:12a}.

We may take the derivative 
\[
\frac{dS}{dt^{\alpha}}=\frac{\partial S}{\partial t^{\alpha}}+\frac{\partial S}{\partial q^{a}}\frac{dq^{a}}{dt^{\alpha}}=\pi_{\alpha}+p_{a}\frac{dq^{a}}{dt^{\alpha}}\equiv p_{i'}\frac{dq^{i'}}{dt^{\alpha}},\,\,\,\,\,\,\,\,\left\{ i'\right\} =\left\{ 0,1,\cdots,N\right\} .
\]
The second derivative results 
\[
\frac{d^{2}S}{dt^{\alpha}dt^{\beta}}-p_{i'}\frac{d^{2}q^{i'}}{dt^{\alpha}dt^{\beta}}=\frac{dq^{i'}}{dt^{\alpha}}\frac{dp_{i'}}{dt^{\beta}}.
\]
The left hand side is symmetric in $\alpha$ and $\beta$, so the
skew-symmetric part of the right hand side must be zero. This yields
the condition 
\begin{equation}
\left(t^{\alpha},t^{\beta}\right)=0,\label{eq:17}
\end{equation}
where we define the Lagrange brackets on the complete phase space:
\begin{equation}
\left(t^{\alpha},t^{\beta}\right)\equiv\frac{dq^{i'}}{dt^{\alpha}}\frac{dp_{i'}}{dt^{\beta}}-\frac{dq^{i'}}{dt^{\beta}}\frac{dp_{i'}}{dt^{\alpha}}.\label{eq:18}
\end{equation}
Therefore, the conditions \eqref{eq:17} are necessary for the existence
of a complete solution of the HJ equations.

To show that \eqref{eq:17} are also sufficient, let us suppose a
set of functions 
\[
q^{i'}=q^{i'}\left(t^{\alpha}\right),\,\,\,\,\,\,\,\,\,\,\, p_{i'}=p_{i'}\left(t^{\alpha}\right),
\]
that obeys \eqref{eq:17}. Taking total derivatives of \eqref{eq:17}
we have 
\[
\frac{d}{dt^{\beta}}\left[\frac{dq^{i'}}{dt^{\alpha}}p_{i'}\right]=\frac{d}{dt^{\alpha}}\left[\frac{dq^{i'}}{dt^{\beta}}p_{i'}\right].
\]
Observing the above expression, there must be a function $S\left[q^{i'}\left(t^{\alpha}\right)\right]$
such that 
\begin{equation}
\frac{dq^{i'}}{dt^{\alpha}}p_{i'}=\frac{dS}{dt^{\alpha}}.\label{eq:19}
\end{equation}
In this case, derivation of $S$ yields 
\begin{equation}
\frac{dS}{dt^{\alpha}}=\frac{\partial S}{\partial q^{i'}}\frac{dq^{i'}}{dt^{\alpha}},\label{eq:20}
\end{equation}
and comparing \eqref{eq:19} and \eqref{eq:20}, 
\begin{equation}
p_{i'}=\frac{\partial S}{\partial q^{i'}}\,\,\,\,\,\,\implies\,\,\,\,\,\,\,\,\,\,\,\,\, p_{a}=\frac{\partial S}{\partial q^{a}},\,\,\,\,\,\pi_{\alpha}=\frac{\partial S}{\partial t^{\alpha}}.\label{eq:21}
\end{equation}
Therefore, \eqref{eq:17} is the necessary and sufficient condition
for the existence of a function $S\left(t^{\alpha},q^{a}\right)$
whose gradient follows the direction of the conjugate momenta.

If we take the derivative 
\[
\frac{d}{dt^{\alpha}}\left[p_{i'}-\frac{\partial S}{\partial q^{i'}}\right]=\frac{dp_{i'}}{dt^{\alpha}}-\frac{\partial^{2}S}{\partial q^{i'}\partial t^{\alpha}}-\frac{\partial^{2}S}{\partial q^{i'}\partial q^{j'}}\frac{dq^{j'}}{dt^{\alpha}}=0,
\]
and use \eqref{eq:12a}, we see that 
\[
\frac{dH'_{\alpha}}{dq^{i'}}=0.
\]
The general solution is given by 
\begin{equation}
H'_{\alpha}\left(t^{\beta},q^{a},\pi_{\beta},p_{a}\right)=\textnormal{constant},\label{eq:22}
\end{equation}
where the constant can be taken to be zero without loss of generality.

Therefore, \eqref{eq:16} is a solution of the HJPDEs, and the conditions
\eqref{eq:17} are the necessary and sufficient conditions for the
existence of a complete solution of these equations. They become our
first version of the integrability conditions.

\subsection{Frobenius' integrability conditions\label{sub:3.2}}

$ $

The conditions \eqref{eq:17} are not very useful, since they demand
knowledge of the solutions of the variational problem. However, using
the CEs \eqref{eq:12a} we may show that 
\[
\left(t^{\alpha},t^{\beta}\right)=\frac{\partial H'_{\alpha}}{\partial t^{\gamma}}\frac{\partial H'_{\beta}}{\partial\pi_{\gamma}}-\frac{\partial H'_{\beta}}{\partial t^{\gamma}}\frac{\partial H'_{\alpha}}{\partial\pi_{\gamma}}+\frac{\partial H'_{\alpha}}{\partial q^{a}}\frac{\partial H'_{\beta}}{\partial p_{a}}-\frac{\partial H'_{\beta}}{\partial q^{a}}\frac{\partial H'_{\alpha}}{\partial p_{a}}=\left\{ H'_{\alpha},H'_{\beta}\right\} .
\]
Therefore, the integrability conditions \eqref{eq:17} can be written
as 
\begin{equation}
\left\{ H'_{\alpha},H'_{\beta}\right\} =0.\label{eq:23}
\end{equation}
The conditions on the Lagrange brackets of the independent variables
becomes conditions on the Poisson brackets of the generators. Eqs.
\eqref{eq:23} are known as the Frobenius' integrability conditions
(FICs).

Note that 
\[
dH'_{\alpha}=\left\{ H'_{\alpha},H'_{\beta}\right\} dt^{\beta},
\]
and, if $t^{\alpha}$ are independent parameters, the FICs imply 
\begin{equation}
dH'_{\alpha}=0.\label{eq:24}
\end{equation}
The conditions \eqref{eq:24} and \eqref{eq:23} are completely equivalent,
but the later is more convenient to analyze systems that are not integrable
at first sight. As shown in \cite{GB}, application of \eqref{eq:24}
may reveal dependence of the independent variables, in the form of
total differential equations, leading naturally to the introduction
of generalized brackets. Moreover, \eqref{eq:24} states that the
generators $H'_{\alpha}$ are also a set of dynamical invariants.

We may generalize \eqref{eq:23} to 
\begin{equation}
\left\{ H'_{\alpha},H'_{\beta}\right\} =C_{\alpha\beta}^{\,\,\,\,\,\,\gamma}H'_{\gamma}.\label{eq:25}
\end{equation}
The proof for the case of classical mechanics can be found in \cite{Fom},
but involves a highly mathematical labor. However, we can easily convince
ourselves that \eqref{eq:25} is a proper generalization. First, we
notice that the meaning of \eqref{eq:24} is that the generators must
be dynamical invariants. Second, we may use the Jacobi identity to
show that the PBs of two dynamical invariants is another dynamical
invariant. If the set $H'_{\alpha}$ closes the Poisson algebra \eqref{eq:25},
it means that this set is a complete set of invariants. In this case,
we clearly have preserved the relations \eqref{eq:24} in the reduced
phase space, where $H'_{\alpha}=0$, since \eqref{eq:25} implies
\[
dH'_{\alpha}=\left\{ H'_{\alpha},H'_{\beta}\right\} dt^{\beta}=C_{\alpha\beta}^{\,\,\,\,\,\,\gamma}H'_{\gamma}dt^{\beta}=0.
\]
Therefore, even if \eqref{eq:25} are obeyed instead of the stronger
conditions \eqref{eq:23}, \eqref{eq:24} still hold in $\mathbf{T}^{*}\mathbb{Q}_{P}$,
and the dynamics in this reduced phase space is independent for each
parameter $t^{\alpha}$.

\section{Canonical transformations generated by a complete set of involutive
constraints\label{sec:4}}

$ $

In regular classical mechanics, temporal evolution can be seen as
a set of successive infinitesimal canonical transformations \cite{Textbooks}.
This becomes evident since we may write solutions of Hamilton's equations
as hamiltonian flows generated by the hamiltonian function, and these
flows are canonical in the sense that they preserve the volume of
the phase space (Liouville theorem). For constrained systems we may
show that this picture is also valid. In order to build this picture
we now turn to the geometrical aspects of the HJ formalism.

\subsection{The geometric approach\label{sub:4.1}}

$ $

Let us define the set of vector fields 
\begin{equation}
X_{\alpha}\equiv\frac{d}{dt^{\alpha}}=\chi_{\alpha}^{I}\,\frac{\partial}{\partial\xi^{I}},\,\,\,\,\,\,\,\,\,\,\,\,\,\chi_{\alpha}^{I}\equiv\left\{ \xi^{I},H'_{\alpha}\right\} ,\label{eq:26}
\end{equation}
where $\xi^{I}=\left(t^{\alpha},q^{a},\pi_{\alpha},p_{a}\right)$.
With these vector fields, the fundamental differential \eqref{eq:14}
can be written as 
\begin{equation}
dF=dt^{\alpha}X_{\alpha}F.\label{eq:27}
\end{equation}
We may also rewrite the CEs using \eqref{eq:27}. They have the form
of the TDEs 
\begin{equation}
d\xi^{I}=dt^{\alpha}X_{\alpha}\xi^{I},\,\,\,\,\,\,\,\,\,\,\,\,\, dS=dt^{\alpha}X_{\alpha}S.\label{eq:28}
\end{equation}

Now we compute the Lie derivative between two of these vector fields:
\[
\mathcal{L}_{X^{\alpha}}X_{\beta}F=\left[X_{\alpha},X_{\beta}\right]F=\left\{ \left\{ F,H'_{\beta}\right\} ,H'_{\alpha}\right\} -\left\{ \left\{ F,H'_{\alpha}\right\} ,H'_{\beta}\right\} .
\]
Applying the Jacobi identity on the right hand side, also considering
\eqref{eq:25}, we have 
\[
\mathcal{L}_{x^{\alpha}}X_{\beta}F=-C_{\alpha\beta}^{\,\,\,\,\,\,\gamma}X_{\gamma}F+\left\{ C_{\alpha\beta}^{\,\,\,\,\,\,\gamma},F\right\} H'_{\gamma}.
\]
If the structure coefficients are independent of $\xi^{I}$, the integrability
conditions become conditions over the Lie brackets between the vector
fields, 
\begin{equation}
\left[X_{\alpha},X_{\beta}\right]=f_{\alpha\beta}^{\,\,\,\,\,\,\gamma}X_{\gamma},\,\,\,\,\,\,\,\,\,\,\,\,\,\,\,\,\, f_{\alpha\beta}^{\,\,\,\,\,\,\gamma}\equiv-C_{\alpha\beta}^{\,\,\,\,\,\,\gamma}.\label{eq:29}
\end{equation}
These are also sufficient conditions for a set of vectors $X_{\alpha}$
to be a complete basis, therefore these vectors span a vector space
of dimension $R+1$. We may identify this vector space with $\Gamma_{R+1}$,
which is the space of the independent variables $t^{\alpha}$.

\subsection{The symplectic structure\label{sub:4.2}}

$ $

We may build the symplectic structure considering the Pfaffian 1-form
defined by $\theta_{c}\equiv dS$. According to the CE \eqref{eq:12b},
\begin{equation}
\theta_{c}=p_{a}dq^{a}+\pi_{\alpha}dt^{\alpha}-H'_{\alpha}dt^{\alpha}.\label{eq:30}
\end{equation}

The symplectic 2-form is defined by $\omega\equiv-d\theta_{c}$, which
gives the expression 
\begin{equation}
\omega=\omega_{P}+a,\label{eq:31}
\end{equation}
where\begin{subequations}\label{eq:32} 
\begin{gather}
\omega_{P}\equiv dq^{a}\wedge dp_{a},\label{eq:32a}\\
a\equiv dt^{\alpha}\wedge d\pi_{\alpha}+dH'_{\alpha}\wedge dt^{\alpha}.\label{eq:32b}
\end{gather}
\end{subequations}

The 2-form $\omega_{P}$ can be written by 
\begin{equation}
\omega_{P}=dq^{a}\wedge dp_{a}=\frac{1}{2}d\xi^{A}\omega_{AB}d\xi^{B},\label{eq:33}
\end{equation}
where $\xi^{A}=\left(q^{a},p_{a}\right)$. The matrix $\omega_{AB}$
is given by 
\begin{equation}
\omega_{AB}\equiv\left(\begin{array}{cc}
0 & \delta_{ab}\\
-\delta_{ab} & 0
\end{array}\right).\label{eq:34}
\end{equation}

On the other hand, since $dH'_{\alpha}=0$, 
\begin{equation}
a=dt^{\alpha}\wedge d\pi_{\alpha}=\left(\frac{\partial^{2}S}{\partial t^{\alpha}\partial t^{\beta}}\right)dt^{\alpha}\wedge dt^{\beta}.\label{eq:35}
\end{equation}
The expression in brackets is symmetric in $\alpha$ and $\beta$,
therefore $a$ is identically zero.

We see that the symplectic structure $\omega$ is singular. Under
the assumption of integrability, it becomes the sum of a regular 2-form
$\omega_{P}$ and a null 2-form $a$. We notice that $\omega_{P}$
is non-degenerate. This is not the case of the full 2-form $\omega$:
if we take the vectors $X_{\alpha}$ we get the contraction 
\begin{equation}
dH'_{\alpha}=i_{X_{\alpha}}\omega=0,\label{eq:36}
\end{equation}
if \eqref{eq:24} hold. Therefore, we recover the geometric nature
of the singularity of a given system: it comes from the fact that
the symplectic structure is degenerate, and the vector fields $X_{\alpha}$
are the eigenvectors that correspond to its null space. We also have
\begin{equation}
i_{X_{\alpha}}i_{X_{\beta}}\omega=\left\{ H'_{\alpha},H'_{\beta}\right\} ,\label{eq:37}
\end{equation}
so the FICs \eqref{eq:23} can also be written as $i_{X_{\alpha}}i_{X_{\beta}}\omega=0$.

\subsection{Canonical transformations and characteristic flows\label{sub:4.3}}

$ $

Now let us see how the vector fields $X_{\alpha}$ generates active
canonical transformations in $\mathbf{T}^{*}\mathbb{Q}_{N+1}$. The
general form of these transformations is naturally given by the structure
of the fundamental differential \eqref{eq:27}. Let us define an infinitesimal
transformation $\delta t^{\alpha}\equiv\bar{t}^{\alpha}-t^{\alpha}$
on the independent variables. In principle, $\delta t^{\alpha}$ are
arbitrary (small) functions of $\xi^{I}$. In this case, it implies
the transformation 
\begin{equation}
\delta F=\delta t^{\alpha}X_{\alpha}F,\label{eq:38}
\end{equation}
for any function $F\left(\xi^{I}\right)$ of $\mathbf{T}^{*}\mathbb{Q}_{N+1}$.
We remark that this transformation is not generally related to the
dynamics, so the characteristics equations and the HJ equations $H'_{\alpha}=0$
may not be satisfied. However, if we choose $\delta t^{\alpha}=dt^{\alpha}$,
\eqref{eq:38} becomes the fundamental differential \eqref{eq:27}
of the system. Therefore, the dynamical evolution becomes a special
case of the transformation \eqref{eq:38}.

If $F=\xi^{I}$, \eqref{eq:38} defines transformations in the coordinates
of $\mathbf{T}^{*}\mathbb{Q}_{N+1}$. We may write these transformations
as 
\begin{equation}
\xi^{I}\left(\bar{t}^{\alpha}\right)=g\xi^{I}\left(t^{\alpha}\right),\label{eq:39}
\end{equation}
where we define the operator 
\begin{equation}
g\equiv1+\delta t^{\alpha}X_{\alpha}.\label{eq:40}
\end{equation}
In this case we say that $g$ carries the infinitesimal flows generated
by $X_{\alpha}$. Let us call these flows the characteristic flows
(CFs) of the system.

The CFs are active canonical transformations. We may see this by taking
the application $g\omega g^{-1}$, where $g^{-1}\equiv1-\delta t^{\alpha}X_{\alpha}$
is the inverse transformation. Supposing integrability, we have 
\begin{equation}
gg^{-1}=g^{-1}g=1,\label{eq:41}
\end{equation}
and then, 
\begin{equation}
g\omega g^{-1}=\omega-\delta t^{\alpha}\delta t^{\beta}i_{X_{\alpha}}i_{X_{\beta}}\omega=\omega.\label{eq:42}
\end{equation}
So $\omega$ is preserved by \eqref{eq:38}. Of course, invariance
of the symplectic 2-form $\omega$ is reflected in any $2p$-form
\[
\omega^{\wedge p}\equiv\underbrace{\omega\wedge\omega\wedge\cdots\wedge\omega}_{p},
\]
specially the volume $2\left(N+1\right)-$form $v\approx\omega^{\wedge2\left(N+1\right)}$,
whose invariance is known as the Liouville theorem. Because $\omega=\omega_{P}+a$,
and $a$ is a null-form, the volume element $a^{\wedge2\left(R+1\right)}$
is identically zero, and all the above properties are also applied
to $\omega_{P}$.

Now, suppose a 2-dimensional surface $\Lambda\subset\mathbf{T}^{*}\mathbb{Q}_{N+1}$.
The area of this surface is calculated by 
\[
I_{\Lambda}=\int_{\Lambda}\omega=-\int_{\Lambda}d\theta_{c}=-\oint_{\partial\Lambda}\theta_{c}.
\]
Then, invariance of $\omega$ implies that $I_{\Lambda}$ is preserved
by the CFs. On the right side we have the integral of the canonical
1-form $\theta_{c}$ over a closed curve $\partial\Lambda$. If this
integral is preserved, the integral 
\begin{equation}
S=\int_{C}\theta_{c}\label{eq:43}
\end{equation}
is path independent. This integral defines a canonical fundamental
integral in $\mathbf{T}^{*}\mathbb{Q}_{P}$, canonical action that
was already found in the form of the total differential equation \eqref{eq:12b}.
This action is, then, invariant under the transformations \eqref{eq:38}
apart of boundary terms.

Let us suppose the case in which the FICs \eqref{eq:25} imply 
\begin{equation}
\left[X_{\alpha},X_{\beta}\right]=f_{\alpha\beta}^{\,\,\,\,\,\,\gamma}X_{\gamma},\label{eq:44}
\end{equation}
which happens to be the necessary and sufficient conditions for $X_{\alpha}$
to be a complete basis of $\Gamma_{R+1}$. Supposing a function $F\in\mathbf{T}^{*}\mathbb{Q}_{N+1}$,
we may build the composition of two flows $g_{\epsilon}$ and $g_{\lambda}$
\[
g_{\epsilon}\equiv1+\epsilon^{\alpha}X_{\alpha},\,\,\,\,\,\,\,\,\, g_{\lambda}\equiv1+\lambda^{\alpha}X_{\alpha}.
\]
This composition yields the Lie bracket 
\[
\left[g_{\epsilon},g_{\lambda}\right]F=g_{\upsilon}F,\,\,\,\,\,\,\, g_{\upsilon}\equiv1+\upsilon^{\alpha}X_{\alpha},\,\,\,\,\,\,\upsilon^{\gamma}\equiv\epsilon^{\alpha}\lambda^{\beta}f_{\alpha\beta}^{\,\,\,\,\,\gamma}.
\]

Therefore, the composition of two characteristic flows is another
characteristic flow. This is the group property that allows the finite
composition 
\begin{equation}
g\left(\Delta t^{\alpha}\right)=\exp\left[\Delta t^{\alpha}X_{\alpha}\right],\label{eq:45}
\end{equation}
to become an element of a Lie group of canonical transformations.
In other words, if the algebra of the involutive constraints is reflected
on the vector fields $X_{\alpha}$, we may build a Lie group out of
the Lie algebra of those vector fields.

As result, we reach our first objective, which is to show that a complete
set of involutive constraints $H'_{\alpha}=0$ are generators of infinitesimal
canonical transformations with the form 
\begin{equation}
\delta\xi^{I}=\left\{ \xi^{I},H'_{\alpha}\right\} \delta t^{\alpha},\label{eq:46}
\end{equation}
which are called the characteristic flows of the system.

\subsection{Connection to gauge transformations\label{sub:4.4}}

$ $

Among the CFs \eqref{eq:46}, we may define a special class of transformations.
We set $\delta t^{0}=\delta t=0$, which means that these transformations
are taken at constant time. They are given by 
\begin{equation}
\delta\xi^{I}=\left\{ \xi^{I},H'_{z}\right\} \delta t^{z}.\label{eq:47}
\end{equation}
Although \eqref{eq:47} has the same structure of the canonical flows,
this is not sufficient to assure canonicity. The generators must be
in involution among themselves, 
\begin{equation}
\left\{ H'_{z},H'_{x}\right\} =C_{zx}^{\,\,\,\,\,\, y}H'_{y}.\label{eq:48}
\end{equation}

Remember that the algebra that assures integrability is the complete
Poisson algebra given by \eqref{eq:25}, which includes the generator
of time displacement $H'_{0}$. The components $\left(z,x\right)$
of that expression are given by 
\[
\left\{ H'_{z},H'_{x}\right\} =C_{zx}^{\,\,\,\,\, y}H'_{y}+C_{zx}^{\,\,\,\,\,0}H'_{0},
\]
so \eqref{eq:48} does not hold unless the structure coefficients
$C_{zx}^{\,\,\,\,\,0}$ are zero, or the constraints $H'_{\alpha}=0$
are valid. If we impose $C_{zx}^{\,\,\,\,\,0}=0$, it is implied that
the bracket $\left\{ H'_{z},H'_{0}\right\} $ must be identically
zero in the complete phase space $\mathbf{T}^{*}\mathbb{Q}_{N+1}$.
This restriction is simply too strong and cannot be accomplished in
general. On the other hand, we may demand that the HJ equations $H'_{\alpha}=0$
are valid. In this case, the algebrae \eqref{eq:25} and \eqref{eq:48}
become abelian, and the transformations \eqref{eq:47} become restricted
to $\mathbf{T}^{*}\mathbb{Q}_{P}$.

Therefore, along with the condition that $\delta\xi^{I}$ are taken
at constant $t$, we may impose that they cannot leave the reduced
phase space. In this case, \eqref{eq:47} become the same transformations
that was called \textquotedbl{}point transformations\textquotedbl{}
by Dirac \cite{Dirac1}, in the hamiltonian picture.

The point transformations \eqref{eq:47} are generated by the function
\begin{equation}
G=H'_{z}\delta t^{z},\label{eq:49}
\end{equation}
since 
\begin{equation}
\delta_{G}\xi^{I}=\left\{ \xi^{I},G\right\} =\left\{ \xi^{I},H'_{z}\right\} \delta t^{z},\label{eq:50}
\end{equation}
when $H'_{z}=0$. Therefore $\delta_{G}\xi^{I}$ becomes equal to
\eqref{eq:47} in $\mathbf{T}^{*}\mathbb{Q}_{P}$, and $G$ is called
their generating function. These transformations are canonical symmetries,
since the algebra \eqref{eq:48} assures that the symplectic structure
is not changed by \eqref{eq:50}.

On the other hand, gauge transformations are Noether symmetries of
the fundamental integral. Suppose a set of infinitesimal transformations
$\delta t=\bar{t}-t$ and $\delta q^{i}=\bar{q}^{i}-q^{i}$, the change
in the action \eqref{eq:01} is given by 
\begin{equation}
\delta I=\int dt\left[\left(\frac{\partial L}{\partial q^{i}}-\frac{d}{dt}\frac{\partial L}{\partial\dot{q}^{i}}\right)\left(\delta q^{i}-\delta t\frac{dq^{i}}{dt}\right)+\frac{d}{dt}\left(\frac{\partial L}{\partial\dot{q}^{i}}\delta q^{i}-H\delta t\right)\right],\label{eq:51}
\end{equation}
where 
\begin{equation}
H\equiv\frac{\partial L}{\partial\dot{q}^{i}}\dot{q}^{i}-L\label{eq:52-1}
\end{equation}
is the hamiltonian function in terms of coordinates and velocities.
Eq. \eqref{eq:51} provides the equations of motion 
\begin{equation}
\frac{\partial L}{\partial q^{i}}-\frac{d}{dt}\frac{\partial L}{\partial\dot{q}^{i}}=0,
\end{equation}
and the conjugated momenta $p_{i}=\partial L/\partial\dot{q}^{i}$
of the system, when Hamilton's principle is applied.

To be lagrangian symmetries, however, the transformations \eqref{eq:50},
with $\delta t=0$, must obey the Lie equation \cite{Mukunda}

\begin{equation}
\delta L=\left(\frac{\partial L}{\partial q^{i}}-\frac{d}{dt}\frac{\partial L}{\partial\dot{q}^{i}}\right)\delta q^{i}+\frac{d}{dt}\left(\frac{\partial L}{\partial\dot{q}^{i}}\delta q^{i}\right)=0,\label{eq:54-1}
\end{equation}
for the coordinate transformations $\delta q^{i}=\left\{ q^{i},H'_{z}\right\} \delta t^{z}=\chi_{z}^{i}\delta t^{z}$.
In the following applications, we will see that \eqref{eq:54-1} implies
linear dependency between some of the variations $\delta t^{z}$.
The generator of the CFs \eqref{eq:49}, then, becomes the generator
of the point transformations, usually up to a total time derivative.
For field theories, \eqref{eq:51}, \eqref{eq:52-1} and \eqref{eq:54-1}
have straightforward generalizations.

\section{Applications\label{sec:5}}

$ $

In this section we apply the ideas discussed in the past sections
in specific examples. The main script is the following: first we find
the complete set of involutive constraints of the theory, therefore
building its integrability. If necessary, non-involutive constraints
are treated by the method developed in \cite{GB}, and all Poisson
brackets must be changed to generalized brackets. We then build the
CEs of the theory, and proceed to the analysis of their characteristic
flows and related gauge transformations.

\subsection{The Christ-Lee model\label{sec:5.1}}

$ $

We begin with the Christ-Lee model \cite{Christ-Lee}. It can be considered
a toy model in classical mechanics, but unlike the usual toy models
in the literature, the Christ-Lee system is a very conceivable mechanical
system, although a very special one. It is simply a particle on a
plane, whose position is given by a vector $\mathbf{x}=\left(x_{1},x_{2}\right)$,
submitted to the constraint that its position and momentum lie in
the same direction.

The lagrangian of the Christ-Lee model is given by 
\begin{equation}
L\left(\mathbf{x},\dot{\mathbf{x}},q\right)=\frac{1}{2}\,\dot{\mathbf{x}}^{2}-q\,\mathbf{x}\cdot\varepsilon\cdot\dot{\mathbf{x}}+\frac{1}{2}q^{2}\mathbf{x}^{2}-V\left(\mathbf{x}^{2}\right).\label{eq:52}
\end{equation}
The matrix $\varepsilon$ is the skew-symmetric matrix 
\begin{equation}
\varepsilon=\left(\begin{array}{cc}
0 & 1\\
-1 & 0
\end{array}\right).\label{eq:53}
\end{equation}
In this, and in the next example, we use the dot $\cdot$ to denote
a scalar product. In this case $\dot{\mathbf{x}}^{2}$ represents
$x_{i}x^{i}$, $\mathbf{x}\cdot\varepsilon\cdot\dot{\mathbf{x}}$
actually means $\varepsilon_{ij}x^{i}\dot{x}^{j}$, and $\varepsilon\cdot\mathbf{x}$
states for $\varepsilon_{ij}x^{i}=-\varepsilon_{ji}x^{i}$, where
$i,j=1,2$.

The Euler-Lagrange equations of the Christ-Lee model are \begin{subequations}\label{eq:55}
\begin{gather}
\mathbf{x}\cdot\left(q\mathbf{x}-\varepsilon\cdot\dot{\mathbf{x}}\right)=0,\label{eq:55a}\\
\ddot{\mathbf{x}}+2q\,\varepsilon\cdot\dot{\mathbf{x}}+\dot{q}\,\varepsilon\cdot\mathbf{x}-q^{2}\mathbf{x}=-\nabla V,\label{eq:55b}
\end{gather}
\end{subequations}and the momenta 
\begin{equation}
\mathbf{p}\equiv\dot{\mathbf{x}}-q\,\mathbf{x}\cdot\varepsilon,\,\,\,\,\,\,\,\,\,\,\,\,\,\pi=0,\label{eq:56}
\end{equation}
conjugated with the variables $\mathbf{x}$ and $q$, respectively.
The definition for $\mathbf{p}$ gives equations for the velocities
\begin{equation}
\dot{\mathbf{x}}=\mathbf{p}+q\,\mathbf{x}\cdot\varepsilon,\label{eq:57}
\end{equation}
and the last momentum is a canonical constraint 
\begin{equation}
H'_{1}\equiv\pi=0.\label{eq:58}
\end{equation}

The canonical hamiltonian is given by 
\begin{equation}
H_{0}=\frac{1}{2}\,\mathbf{p}^{2}+V+q\,\mathbb{L},\label{eq:59}
\end{equation}
where $\mathbb{L}\equiv\mathbf{x}\cdot\varepsilon\cdot\mathbf{p}=x_{1}p_{2}-x_{2}p_{1}$
is the angular momentum in two dimensions. In this case we have two
HJ equations:\begin{subequations}\label{eq:60} 
\begin{gather}
H'_{0}\equiv\pi_{0}+H_{0}=0,\label{eq:60a}\\
H'_{1}\equiv\pi=0.\label{eq:60b}
\end{gather}
\end{subequations}Using the fundamental PBs $\left\{ \mathbf{x},\mathbf{p}\right\} =\mathbf{1}$
and $\left\{ q,\pi\right\} =1$, where $\mathbf{1}$ is the identity
in two dimensions, we verify that the PB between the \eqref{eq:60}
are $\left\{ H'_{1},H'_{1}\right\} =0,$ and $\left\{ H'_{1},H'_{0}\right\} =-\mathbb{L}.$
Then the system of HJ equations is not integrable, requiring the imposition
of another constraint, 
\begin{equation}
H'_{2}\equiv\mathbb{L}=0.\label{eq:61}
\end{equation}
It is straightforward to show that the system \eqref{eq:60} and \eqref{eq:61}
is completely integrable. The final algebra is given by 
\begin{equation}
\left\{ H'_{1},H'_{0}\right\} =-H'_{2},\label{eq:62}
\end{equation}
with all other PBs identically zero.

The fundamental differential of an observable $F$ is given by 
\begin{equation}
dF=\left\{ F,H'_{\alpha}\right\} dt^{\alpha}=\left\{ F,H'_{0}\right\} dt^{0}+\left\{ F,H'_{1}\right\} dt^{1}+\left\{ F,H'_{2}\right\} dt^{2},\label{eq:63}
\end{equation}
where $t^{\alpha}=\left(t^{0}=t,t^{1}=q,t^{2}\right)$ are the independent
variables. Because $H'_{2}$ comes from the integrability conditions,
we must expand the parameter space, and therefore the complete phase
space, with a new parameter $t^{2}$. This is done so that $H'_{2}$
becomes a generator of the dynamics in the direction of $t^{2}$.

The characteristic equations are given by\begin{subequations}\label{eq:64}
\begin{gather}
dq=dt^{1}\label{eq:64a}\\
d\pi=-H'_{2}dt,\label{eq:64b}\\
d\mathbf{x}=\left[\mathbf{p}-q\,\varepsilon\cdot\mathbf{x}\right]dt+\varepsilon\cdot\mathbf{x}dt^{2},\label{eq:64c}\\
d\mathbf{p}=-\left[q\,\varepsilon\cdot\mathbf{p}+\nabla V\right]dt+\varepsilon\cdot\mathbf{p}dt^{2}.\label{eq:64d}
\end{gather}
\end{subequations}Since integrability is assured, $t^{2}$ and $t$
are LI, then the time evolution gives the set of equations\begin{subequations}\label{eq:65}
\begin{gather}
\dot{q}=0\label{eq:65a}\\
\dot{\pi}=-H'_{2}=0,\label{eq:65b}\\
\dot{\mathbf{x}}=\mathbf{p}-q\,\varepsilon\cdot\mathbf{x},\label{eq:65c}\\
\dot{\mathbf{p}}=-q\,\varepsilon\cdot\mathbf{p}-\nabla V.\label{eq:65d}
\end{gather}
\end{subequations}

Relation \eqref{eq:65c} reproduces \eqref{eq:57}. Taking the derivative
in $t$, substituting \eqref{eq:65d} and using \eqref{eq:65c} again,
we have the Euler-Lagrange equations \eqref{eq:55b}. Then, equivalence
is actually assured between the time evolution of the CEs and the
Euler-Lagrange equations.

Apart of the time evolution, the generators $H'_{1}$ and $H'_{2}$
give the canonical transformations $\delta F=\left\{ F,H'_{z}\right\} \delta t^{z}$,
for $z=1,2$. It results in 
\begin{gather}
\delta q=\delta t^{1},\,\,\,\,\,\,\,\,\,\,\,\,\,\,\,\,\,\,\,\,\delta\mathbf{x}=\varepsilon\cdot\mathbf{x}\delta t^{2},\,\,\,\,\,\,\,\,\,\,\,\,\,\,\,\,\,\,\,\,\delta\mathbf{p}=\varepsilon\cdot\mathbf{p}\delta t^{2}.\label{eq:66}
\end{gather}
The transformation on $q$ is just arbitrary, it depends of the form
of $\delta t^{1}$. On the other hand, the CT on $\mathbf{x}$ and
$\mathbf{p}$ are infinitesimal rotations. The generator of the characteristic
flows in the directions of $t^{1}$ and $t^{2}$ is given by 
\begin{equation}
G_{CF}=H'_{z}\delta t^{z}=H'_{1}\delta t^{1}+H'_{2}\delta t^{2}.\label{eq:67}
\end{equation}
As we saw earlier, the algebra of the generators $H'_{1}$ and $H'_{2}$
is abelian.

Now, the first variation of \eqref{eq:52} under infinitesimal transformations
$\delta\mathbf{x}\equiv\mathbf{x}'-\mathbf{x}$ and $\delta q=q'-q$
is given by 
\begin{eqnarray}
\delta L & = & \frac{d}{dt}\left[\left(\dot{\mathbf{x}}-q\,\mathbf{x}\cdot\varepsilon\right)\cdot\delta\mathbf{x}\right]\nonumber \\
 &  & -\delta\mathbf{x}\cdot\left[\ddot{\mathbf{x}}+2q\,\varepsilon\cdot\dot{\mathbf{x}}+\dot{q}\,\varepsilon\cdot\mathbf{x}-q^{2}\mathbf{x}+\nabla V\right]\nonumber \\
 &  & +\delta q\,\mathbf{x}\cdot\left(q\mathbf{x}-\varepsilon\cdot\dot{\mathbf{x}}\right).\label{eq:54}
\end{eqnarray}
If \eqref{eq:66} are symmetries of the lagrangian function, $\delta L$
must be zero. Then, substituting \eqref{eq:66} in \eqref{eq:54},
and considering $V=V\left(\mathbf{x}^{2}\right)$, we have
\begin{equation}
\delta L=\left(\frac{d}{dt}\delta t^{2}+\delta t^{1}\right)\left(q\mathbf{x}^{2}-\mathbf{x}\cdot\varepsilon\cdot\dot{\mathbf{x}}\right).\label{eq:68}
\end{equation}
In this case $\delta L=0$ if $\delta t^{1}=-d\left(\delta t^{2}\right)/dt$,
independently of the Euler-Lagrange equations. Let us suppose that
$\delta t^{2}=\theta\left(t\right)$. In this case, $\delta t^{1}=-\dot{\theta}$,
and 
\begin{equation}
G_{g}=-H'_{1}\dot{\theta}-H'_{2}\theta=-\left(\pi\dot{\theta}+\mathbb{L}\theta\right)\label{eq:69}
\end{equation}
is the generator of the gauge transformations\begin{subequations}\label{eq:70}
\begin{gather}
\delta q=\left\{ q,G\right\} =-\dot{\theta},\label{eq:70a}\\
\delta\mathbf{x}=\left\{ \mathbf{x},G\right\} =-\mathbf{x}\cdot\varepsilon\cdot\left\{ \mathbf{x},\mathbf{p}\right\} \theta=\varepsilon\cdot\mathbf{x}\,\theta.\label{eq:70b}
\end{gather}
\end{subequations}

\subsection{Chern-Simons quantum mechanics\label{sub:5.2}}

$ $

Now let us consider the two dimensional movement of a charged particle
in a constant magnetic field $B$, and a quadratic scalar potential.
This system is described by the Lagrange function 
\[
L\left(\mathbf{x},\dot{\mathbf{x}}\right)=\frac{1}{2}\, m\dot{\mathbf{x}}^{2}+\frac{B}{2}\,\mathbf{x}\cdot\varepsilon\cdot\dot{\mathbf{x}}-\frac{k}{2}\,\mathbf{x}^{2},
\]
and it is known to be the mechanical analogous of the three-dimensional
topologically massive electrodynamics in the Weyl gauge. The term
of the magnetic field actually corresponds to a pure Chern-Simons
term in the three-dimensional gauge theory. In the limit $m,k\rightarrow0$
the quantization of this model results in a quantum mechanical theory
with interesting topological effects \cite{Jackiw}.

This model can be made more interesting with the inclusion of another
Chern-Simons term, so we will work with the system described by the
function 
\begin{equation}
L\left(\mathbf{x},\dot{\mathbf{x}},q\right)=\frac{B}{2}\,\mathbf{x}\cdot\varepsilon\cdot D\mathbf{x}+\nu q,\,\,\,\,\,\,\,\,\,\, D\mathbf{x}\equiv\dot{\mathbf{x}}+q\varepsilon\cdot\mathbf{x}.\label{eq:71}
\end{equation}
As in the past example, $\mathbf{x}$ is a position vector in two
dimensional euclidian space, $q$ is an auxiliary scalar variable
and $\nu$ is a numerical parameter. The matrix $\varepsilon$ is
the same defined in \eqref{eq:53}.

The equations of motion of the Chern-Simons quantum mechanics are\begin{subequations}\label{eq:73}
\begin{gather}
\nu=\frac{B}{2}\,\mathbf{x}^{2},\label{eq:73a}\\
D\mathbf{x}=0,\label{eq:73b}
\end{gather}
\end{subequations}and the conjugate momenta of the variables $q$
and $\mathbf{x}$, 
\begin{equation}
\pi_{q}=0,\,\,\,\,\,\,\,\,\,\,\,\,\,\,\,\,\,\,\,\mathbf{p}=-\frac{B}{2}\,\varepsilon\cdot\mathbf{x},\label{eq:74}
\end{equation}
respectively. The canonical hamiltonian takes the form 
\begin{equation}
H_{0}=-q\left(\nu-\frac{B}{2}\,\mathbf{x}^{2}\right).\label{eq:75}
\end{equation}
Therefore, we have the following set of HJPDEs:\begin{subequations}\label{eq:76}
\begin{gather}
H'_{0}\equiv\pi_{t}+H_{0}=0,\label{eq:76a}\\
H'_{1}\equiv\pi_{q}=0,\label{eq:76b}\\
\boldsymbol{\psi}\equiv\mathbf{p}+\frac{B}{2}\,\varepsilon\cdot\mathbf{x}=0.\label{eq:76c}
\end{gather}
\end{subequations}We notice that the last constraint is a 2-vector.

This system is not integrable: here we have an example where non involutive
constraints are present. This can be seen by the PB\begin{subequations}\label{eq:77}
\begin{gather}
\left\{ \boldsymbol{\psi},\boldsymbol{\psi}\right\} =B\varepsilon,\label{eq:77a}\\
\left\{ \boldsymbol{\psi},H'_{0}\right\} =qB\mathbf{x},\label{eq:77b}\\
\left\{ H'_{1},H'_{0}\right\} =\nu-\frac{B}{2}\,\mathbf{x}^{2}.\label{eq:77c}
\end{gather}
\end{subequations}Following the procedure outlined in \cite{GB},
\eqref{eq:77a} indicates that we may introduce the GB 
\begin{equation}
\left\{ F,G\right\} ^{*}=\left\{ F,G\right\} -\frac{1}{B}\left\{ F,\boldsymbol{\psi}\right\} \cdot\varepsilon\cdot\left\{ \boldsymbol{\psi},G\right\} ,\label{eq:78}
\end{equation}
which define the fundamental relations 
\begin{equation}
\left\{ q,\pi_{q}\right\} ^{*}=1,\,\,\,\,\,\,\,\,\left\{ \mathbf{x},\mathbf{x}\right\} ^{*}=\frac{1}{B}\varepsilon,\,\,\,\,\,\,\,\,\left\{ \mathbf{x},\mathbf{p}\right\} ^{*}=\frac{1}{2}\,\mathbf{1},\,\,\,\,\,\,\,\,\left\{ \mathbf{p},\mathbf{p}\right\} ^{*}=-\frac{B}{4}\varepsilon.\label{eq:79}
\end{equation}
Because $\left\{ \boldsymbol{\psi},F\right\} ^{*}=0$ identically
for any $F$, all GB between the constraints are zero, except 
\begin{equation}
\left\{ H'_{1},H'_{0}\right\} ^{*}=\nu-\frac{B}{2}\,\mathbf{x}^{2}.\label{eq:80}
\end{equation}
Therefore, a new HJ equation 
\begin{equation}
H'_{2}\equiv\nu-\frac{B}{2}\,\mathbf{x}^{2}=0\label{eq:81}
\end{equation}
should be added to the system \eqref{eq:76}. It is straightforward
to show that integrability for $H'_{3}$ is obeyed, then the system
is completely integrable with the GB \eqref{eq:78}. Particularly,
we have $\left\{ H'_{2},H'_{1}\right\} ^{*}=0$.

Equations of motion are calculated by the fundamental differential
\begin{equation}
dF=\left\{ F,H'_{\alpha}\right\} ^{*}dt^{\alpha}=\left\{ F,H'_{0}\right\} ^{*}dt^{0}+\left\{ F,H'_{1}\right\} ^{*}dt^{1}+\left\{ F,H'_{2}\right\} ^{*}dt^{2},\label{eq:82}
\end{equation}
where $t^{\alpha}=\left(t^{0}=t,t^{1}=q,t^{2}\right)$ are the independent
variables. As in the preceding example, the generator $H'_{2}$ demands
the expansion of the parameter space with the inclusion of a new independent
variable $t^{2}$. The characteristic equations are given by\begin{subequations}\label{eq:83}
\begin{gather}
dq=dt^{1},\label{eq:83a}\\
d\pi=H'_{2}dt,\label{eq:83b}\\
d\mathbf{x}=-q\varepsilon\cdot\mathbf{x}dt+\varepsilon\cdot\mathbf{x}dt^{2},\label{eq:83c}\\
d\mathbf{p}=\frac{1}{2}Bq\mathbf{x}dt-\frac{1}{2}B\mathbf{x}dt^{2}.\label{eq:83d}
\end{gather}
\end{subequations}The first equation identifies $t^{1}$ with $q$
apart of an arbitrary constant. Eq. \eqref{eq:83b} reproduces the
IC for the involutive constraint $H'_{1}$. This is equivalent to
the constraint $H'_{2}=0$, as expected, and therefore eq. \eqref{eq:73a}
is achieved.

For the remaining equations, we see that time evolution alone gives
\begin{equation}
\dot{\mathbf{x}}=-q\varepsilon\cdot\mathbf{x},\label{eq:84}
\end{equation}
which is the same as the Euler-Lagrange equation \eqref{eq:73b}.
On the other hand, \eqref{eq:83d} becomes 
\begin{equation}
\dot{\mathbf{p}}=\frac{1}{2}Bq\mathbf{x}.\label{eq:85}
\end{equation}
This equation is the time derivative of $\bm{\psi}=0$ when \eqref{eq:84}
is considered. It actually gives Newton's second law for this system.

In addition to this analysis, we write down the canonical transformations
$\delta F=\left\{ F,H'_{z}\right\} \delta t^{z}$, for $z=1,2$, 
\begin{equation}
\delta q=\delta t^{1},\,\,\,\,\,\,\,\,\,\,\,\delta\mathbf{x}=\varepsilon\cdot\mathbf{x}\delta t^{2},\,\,\,\,\,\,\,\,\,\,\,\delta\mathbf{p}=-\frac{1}{2}B\mathbf{x}\delta t^{2}.\label{eq:86}
\end{equation}
The transformation for $q$ is an arbitrary rescaling, and $\delta\mathbf{x}$
is again an infinitesimal rotation. On the other hand, $\delta\mathbf{p}$
is a transformation that mix positions and momenta of the phase space.
It is also straightforward to write the generator of the CFs 
\begin{equation}
G_{CF}=H'_{z}\delta t^{z}=H'_{1}\delta t^{1}+H'_{2}\delta t^{2}.\label{eq:87}
\end{equation}

Now, the first variation of the lagrangian function \eqref{eq:71}
under transformations of the form $\delta\mathbf{x}\equiv\mathbf{x}'-\mathbf{x}$
and $\delta q\equiv q'-q$ is given by
\begin{equation}
\delta L=\frac{d}{dt}\left(\frac{B}{2}\,\mathbf{x}\cdot\varepsilon\cdot\delta\mathbf{x}\right)-B\, D\mathbf{x}\cdot\varepsilon\cdot\delta\mathbf{x}+\left(\nu-\frac{B}{2}\,\mathbf{x}^{2}\right)\delta q.\label{eq:72}
\end{equation}
For the transformations \eqref{eq:86}, the variation \eqref{eq:72}
becomes 
\begin{align}
\delta L & =-\frac{B}{2}\,\mathbf{x}^{2}\left(\frac{d}{dt}\delta t^{2}+\delta t^{1}\right)+\nu\delta t^{1}\nonumber \\
 & =-\frac{B}{2}\,\mathbf{x}^{2}\left[\frac{d}{dt}\delta t^{2}+\delta t^{1}-\delta t^{1}\right]=-\frac{B}{2}\,\mathbf{x}^{2}\frac{d}{dt}\delta t^{2},\label{eq:88}
\end{align}
where \eqref{eq:73a} is used. If $\delta q$ and $\delta\mathbf{x}$
are symmetries of the lagrangian, i.e. $\delta L=0$, we should consider
$\delta t^{2}=\theta$ a time-independent constant. In any case, $\delta L$
is independent of $\delta\omega$, which means that any transformation
in $q$ is a lagrangian symmetry. The generator takes the form 
\begin{equation}
G_{g}=-\pi\delta t^{1}+\left(\frac{B}{2}\,\mathbf{x}^{2}-\nu\right)\theta,\label{eq:89}
\end{equation}
and the gauge transformations are finally given by\begin{subequations}\label{eq:90}
\begin{gather}
\delta q=\left\{ q,G\right\} =\delta t^{1},\label{eq:90a}\\
\delta\mathbf{x}=\left\{ \mathbf{x},G\right\} =\varepsilon\cdot\mathbf{x}\theta.\label{eq:90b}
\end{gather}
\end{subequations}

\subsection{The free Yang-Mills theory\label{sec:5.3}}

$ $

Now let us turn to an example of field theory, the Yang-Mills (YM)
theory without sources, described by the fundamental integral 
\begin{equation}
I\equiv-\frac{1}{4}\int_{\Omega}d\omega F_{\mu\nu}^{a}F^{a\mu\nu}.\label{eq:91}
\end{equation}
In our notation, $\Omega$ is 4-volume in a Minkowski space-time with
metric $\eta=\textnormal{diag}(+---)$, and $d\omega$ is its volume
element. The field strength is defined by 
\begin{equation}
F_{\mu\nu}^{a}\equiv\partial_{\mu}A_{\nu}^{a}-\partial_{\nu}A_{\mu}^{a}-g\, f^{abc}A_{\mu}^{b}A_{\nu}^{c},\label{eq:92}
\end{equation}
where $g$ is a coupling constant and $f^{abc}$ are the structure
coefficients of an $\mathfrak{su}\left(n\right)$ algebra.

Let us define the covariant derivative 
\begin{equation}
D_{\mu}^{ab}\equiv\delta^{ab}\partial_{\mu}-gf^{acb}A_{\mu}^{c}.\label{eq:93}
\end{equation}
The field equations are 
\begin{equation}
D_{\mu}^{ab}F^{b\mu\nu}=0,\label{eq:96}
\end{equation}
and the conjugated covariant momenta are given by 
\begin{equation}
\pi^{a\mu\nu}\equiv F^{a\mu\nu}.\label{eq:97}
\end{equation}
We remark that the gauge transformations related to the fundamental
integral $I$ are 
\begin{equation}
\delta_{g}A_{\mu}^{a}=D_{\mu}^{ab}\epsilon^{b},\label{eq:98}
\end{equation}
where $\epsilon^{a}$ are the gauge parameters. These transformations
leave $I$ invariant up to a boundary term.

To get to the HJ formalism for singular field theories, it is necessary
to choose a particular parametrization for the fields. Here we will
work with the \textquotedbl{}Galilean time\textquotedbl{} $t\equiv x^{0}$,
choice that is known as the instant-form dynamics \cite{rel.dyn.}.
In this case, the conjugated momenta are given by the projection of
\eqref{eq:97} in a unit 4-vector $u\equiv\left(1,0,0,0\right)$:
\begin{equation}
\pi^{a\mu}\equiv\pi^{a\mu\nu}u_{\nu}=\pi^{a\mu0}\equiv-F^{a0\mu}.\label{eq:99}
\end{equation}
For $\mu=0$ it gives 
\begin{equation}
\pi^{a0}=0,\label{eq:100}
\end{equation}
while for $\mu=i$, where $\left\{ i\right\} =\left\{ 1,2,3\right\} $,
\eqref{eq:99} gives an equation for the velocities $\dot{A}_{i}^{a}$,
\begin{equation}
\dot{A}_{i}^{a}=D_{i}^{ab}A_{0}^{b}-\pi_{i}^{a},\label{eq:101}
\end{equation}
where we use the notation $\dot{A}_{\nu}^{a}\equiv\partial_{0}A_{\nu}^{a}$.

We also have the (not symmetric) energy-momentum density: 
\begin{equation}
H_{\mu\nu}\equiv-F_{\mu\gamma}^{a}\partial_{\nu}A_{a}^{\gamma}+\frac{1}{4}\delta_{\mu\nu}F_{\alpha\beta}^{a}F_{a}^{\alpha\beta}.\label{eq:95}
\end{equation}
With \eqref{eq:99}, \eqref{eq:100}, and \eqref{eq:101}, we have
the canonical hamiltonian density $\mathcal{H}_{c}\equiv H_{\mu\nu}u^{\mu}u^{\nu}=H_{00}$,
which up to a total divergence can be written by 
\begin{equation}
\mathcal{H}_{c}=-\frac{1}{2}\pi^{ai}\pi_{i}^{a}-A_{0}^{a}D_{i}^{ab}\pi^{bi}+\frac{1}{4}F_{ij}^{a}F^{aij}.\label{eq:102}
\end{equation}
Therefore the system obeys two HJ equations,\begin{subequations}\label{eq:103}
\begin{gather}
H'_{0}\equiv\pi_{0}+\mathcal{H}_{c}=0,\label{eq:103a}\\
\Phi^{a}\equiv\pi^{a0}=0.\label{eq:103b}
\end{gather}
\end{subequations}

For the theory to be integrable we calculate the PB between \eqref{eq:103},
using the fundamental relations $\left\{ A_{\mu}^{a}\left(x\right),\pi^{b\nu}\left(y\right)\right\} =\delta^{ab}\delta_{\mu}^{\nu}\delta^{3}\left(x-y\right)$.
The only non-zero bracket is given by 
\begin{equation}
\left\{ \Phi^{a},H'_{0}\right\} =D_{i}^{ab}\pi^{bi},\label{eq:104}
\end{equation}
which implies non-integrability. This can be solved by imposing a
new set of constraints 
\begin{equation}
\Gamma^{a}\equiv D_{i}^{ab}\pi^{bi}=0.\label{eq:105}
\end{equation}

Now it is necessary the set \eqref{eq:103}, \eqref{eq:105} to be
in involution. The global sub-algebra of $\Gamma^{a}\left(x\right)$,
calculated with the PB of the variables 
\begin{equation}
\Gamma^{a}\left[h\right]\equiv\int_{\Sigma}d\sigma h\left(x\right)\Gamma^{a}\left(x\right)\label{eq:106}
\end{equation}
is given by 
\begin{equation}
\left\{ \Gamma^{a}\left[h_{1}\right],\Gamma^{b}\left[h_{2}\right]\right\} =-gf^{abc}\Gamma^{c}\left[h_{3}\right],\label{eq:107}
\end{equation}
where $h_{3}\left(x\right)=h_{1}\left(x\right)h_{2}\left(x\right)$,
up to a boundary term in $\partial\Sigma$. In \eqref{eq:106} the
integration is performed in a 3-surface section $\Sigma$ of $\Omega$
at constant $t$, whose volume element is $d\sigma$. The remaining
PB is 
\begin{equation}
\left\{ \Gamma^{a},H'_{0}\right\} =gf^{abc}A_{0}^{b}\Gamma^{c}.\label{eq:108}
\end{equation}
The algebra (\ref{eq:104},\ref{eq:107},\ref{eq:108}) indicates
that the set $\left(H'_{0},\Phi^{a},\Gamma^{a}\right)$ is involutive
with the PBs, therefore integrability is achieved. With the definition
$T^{b}\equiv\left(i/g\right)\Gamma^{a}$, \ref{eq:107} becomes precisely
the $\mathfrak{su}\left(n\right)$ algebra.

Now that we have the complete involutive system of HJ equations for
the YM theory, the characteristic equations can be calculated by the
fundamental differential 
\begin{eqnarray*}
dF & = & \int_{\Sigma}d\sigma_{y}\left[\left\{ F,H'_{0}\left(y\right)\right\} dt+\left\{ F,\Phi^{a}\left(y\right)\right\} \omega^{a}\left(y\right)+\left\{ F,\Gamma^{a}\left(y\right)\right\} d\lambda^{a}\left(y\right)\right]
\end{eqnarray*}
where $\left(t,\omega^{a},\lambda^{a}\right)$ is the set of independent
variables, each one related to its respective generator. Again, because
$\Gamma^{a}$ come from integrability, the new set of independent
variables $\lambda^{a}$ is introduced.

For the variables $A_{\mu}^{a}$ we have 
\begin{equation}
dA_{\mu}^{a}\left(x\right)=\int_{\Sigma}d\sigma_{y}\left\{ A_{\mu}^{a}\left(x\right),H'_{0}\left(y\right)\right\} dt+\delta_{\mu}^{0}d\omega^{a}\left(x\right)-\delta_{\mu}^{i}D_{i}^{ab}\left(x\right)d\lambda^{b}\left(x\right).\label{eq:110}
\end{equation}
Since $\left(t,\omega^{a},\lambda^{a}\right)$ are independent among
themselves, we may write\begin{subequations}\label{eq:111} 
\begin{gather}
\dot{A}_{\mu}^{a}\left(x\right)=\delta_{\mu}^{i}\left[D_{i}^{ab}\left(x\right)A_{0}^{b}\left(x\right)-\pi_{i}^{a}\left(x\right)\right],\label{eq:111a}\\
\frac{\delta A_{\mu}^{a}\left(x\right)}{\delta\omega^{b}\left(y\right)}=\delta^{ab}\delta_{\mu}^{0}\delta^{3}\left(x-y\right),\label{eq:111b}\\
\frac{\delta A_{\mu}^{a}\left(x\right)}{\delta\lambda^{b}\left(y\right)}=-\delta_{\mu}^{i}D_{i}^{ab}\left(x\right)\delta^{3}\left(x-y\right).\label{eq:111c}
\end{gather}
\end{subequations}

Equation \eqref{eq:111b} indicates that $A_{0}^{a}=\omega^{a}$ plus
an arbitrary function independent of the fields, which is a very property
of a degenerate variable in the action. Equation \eqref{eq:111c}
indicates that the dynamics involves the variables $\lambda^{a}$
in the form of the CT $\delta A_{i}^{a}=-D_{i}^{ab}\delta\lambda^{b}$,
which depends on the variables $A_{i}^{a}$. Time evolution is given
by \eqref{eq:111a}, that yields 
\begin{equation}
\dot{A}_{0}^{a}=0,\,\,\,\,\,\,\,\,\,\,\,\,\,\,\,\,\,\dot{A}_{i}^{a}=D_{i}^{ab}A_{0}^{b}-\pi_{i}^{a}.\label{eq:112}
\end{equation}
The equation for $A_{i}^{a}$ reproduces \eqref{eq:101}, as expected.

For the variables $\pi^{a\mu}$ we have 
\begin{equation}
d\pi^{a\mu}\left(x\right)=\int_{\Sigma}d\sigma_{y}\left\{ \pi^{a\mu}\left(x\right),H'_{0}\left(y\right)\right\} dt+gf^{abc}\delta_{i}^{\mu}\pi^{bi}\left(x\right)d\lambda^{c}\left(x\right).\label{eq:113}
\end{equation}
Again, independence of the parameters yields\begin{subequations}\label{eq:114}
\begin{gather}
\dot{\pi}^{a\mu}\left(x\right)=\delta_{0}^{\mu}\Gamma^{a}\left(x\right)-\delta_{i}^{\mu}\left[gf^{abc}A_{0}^{c}\left(x\right)\pi^{bi}\left(x\right)+D_{j}^{ab}\left(x\right)F^{bij}\left(x\right)\right],\label{eq:114a}\\
\frac{\delta\pi^{a\mu}\left(x\right)}{\delta\lambda^{b}\left(y\right)}=-gf^{abc}\delta_{i}^{\mu}\pi^{ci}\left(x\right)\delta^{3}\left(x-y\right).\label{eq:114b}
\end{gather}
\end{subequations}Eq. \eqref{eq:114a} represents time evolution:
\begin{equation}
\dot{\pi}^{a0}=\Gamma^{a},\,\,\,\,\,\,\,\,\,\,\,\,\,\,\,\,\,\,\,\, D_{0}^{ab}\pi^{bi}=-D_{j}^{ab}F^{bij}.\label{eq:115}
\end{equation}
From the first equation, we notice that $\Gamma^{a}=D_{i}^{ab}\pi^{bi}=D_{i}^{ab}F^{bi0}$,
that gives \eqref{eq:96} for $\nu=0$. The second equation is the
field equation \eqref{eq:96} for $\nu=i$, if the use the definition
$\pi^{ai}=F^{ai0}$. Therefore, time evolution is equivalent to the
field equations \eqref{eq:96} for the YM field.

We now turn to the problem of finding the generator of the canonical
transformations defined by equations \eqref{eq:111b}, \eqref{eq:111c}
and \eqref{eq:114b}. They can be written by\begin{subequations}\label{eq:116}
\begin{gather}
\delta A_{\mu}^{a}=\delta_{\mu}^{0}\delta\omega^{a}-\delta_{\mu}^{i}D_{i}^{ab}\delta\lambda^{b},\label{eq:116a}\\
\delta\pi^{a\mu}=gf^{abc}\delta_{i}^{\mu}\pi^{bi}\delta\lambda^{c}.\label{eq:116b}
\end{gather}
\end{subequations}It is straightforward to show that 
\begin{equation}
G_{CT}=\int_{\Sigma}d\sigma\left[\Phi^{a}\delta\omega^{a}+\Gamma^{a}\delta\lambda^{a}\right]\label{eq:117}
\end{equation}
is the generator of the transformations \eqref{eq:116}.

The goal, now, is to find the generator of the gauge transformations.
The first-order variation of the action \eqref{eq:91} under a set
of infinitesimal transformations $\delta x^{\mu}=\bar{x}^{\mu}-x^{\mu}$
and $\delta A_{\mu}^{a}=\bar{A}_{\mu}^{a}-A_{\mu}^{a}$ is given by
\begin{equation}
\delta I=\int_{\Omega}d\omega\left[\partial_{\mu}\left(F^{a\mu\nu}\delta A_{\nu}^{a}-H_{\,\,\nu}^{\mu}\delta x^{\nu}\right)+\left(\delta A_{\nu}^{a}-\delta x^{\gamma}\partial_{\gamma}A_{\nu}^{a}\right)D_{\mu}^{ab}F^{b\mu\nu}\right].\label{eq:94}
\end{equation}
As in the previous examples, the generator can be achieved with \eqref{eq:94},
under the transformations characterized by $\delta x^{\mu}=0$ and
$\delta A_{\mu}^{a}=\delta_{\mu}^{0}\delta\omega^{a}-\delta_{\mu}^{i}D_{i}^{ab}\delta\lambda^{b}$.
After some algebra, and the use of the identities $D_{\mu}^{ab}D_{\nu}^{bc}F^{c\mu\nu}=0$,
the following expression arises: 
\begin{equation}
\delta L=\partial_{\mu}\left[\left(\delta\omega^{a}+D_{0}^{ac}\delta\lambda^{c}\right)F^{a0\mu}\right]+\left(\delta\omega^{a}+D_{0}^{ac}\delta\lambda^{c}\right)D_{\mu}^{ab}F^{b\mu0}.\label{eq:118}
\end{equation}
If the action is invariant under the transformations \eqref{eq:116a},
$\delta L=0$ implies that the correct relationship between the independent
variables of the theory is given by $\delta\omega^{a}=-D_{0}^{ab}\delta\lambda^{b}$.
If we define $\Lambda^{a}\equiv-\delta\lambda^{a}$ as the gauge parameters,
then $\delta\omega^{a}=D_{0}^{ab}\Lambda^{b}$. Of course, the transformation
in $\pi^{a\mu}$ has no analogous in the lagrangian picture. In this
case, the generator \eqref{eq:117} becomes 
\begin{equation}
G_{g}=\int_{\Sigma}d\sigma\left[\Phi^{a}D_{0}^{ab}\Lambda^{b}-\Gamma^{a}\Lambda^{a}\right]=\int_{\Sigma}d\sigma\pi^{a\mu}D_{\mu}^{ab}\Lambda^{b},\label{eq:119}
\end{equation}
up to a boundary term. To check this generator, we calculate 
\begin{equation}
\delta_{g}A_{\mu}^{a}\left(x\right)=\left\{ A_{\mu}^{a}\left(x\right),G_{g}\right\} =D_{\mu}^{ab}\left(x\right)\Lambda^{b}\left(x\right),\label{eq:120}
\end{equation}
which are in fact the gauge transformations \eqref{eq:98} of the
theory. We may perform a further calculation, 
\begin{equation}
\delta_{g}\pi_{\mu}^{a}\left(x\right)=\left\{ \pi_{\mu}^{a}\left(x\right),G_{g}\right\} =-gf^{abc}\pi_{\mu}^{b}\left(x\right)\Lambda^{c}\left(x\right),\label{eq:121}
\end{equation}
which agrees with \eqref{eq:116b}, considering the HJ equation $\pi_{0}^{a}=0$.

\section{Final remarks\label{sec:6}}

$ $As a continuation of the work \cite{GB}, we analyzed the integrability
of constrained systems within the Hamilton-Jacobi formalism, and studied
how this approach links complete sets of involutive HJ equations with
canonical and lagrangian point (gauge) symmetries, as named by Dirac
\cite{Dirac1}, of a fundamental integral. Now let us highlight the
main script of this study.

According to Carathéodory's \textquotedbl{}complete figure\textquotedbl{}
applied to singular systems, the necessary and sufficient condition
for the existence of an extreme configuration of the action \eqref{eq:01}
is the existence of a function $S\left(t^{\alpha},q^{a}\right)$ of
the configuration space that obeys the conditions \eqref{eq:11b},
and is a complete solution of the set of HJ equations \eqref{eq:11a}.
If the independent variables $t^{\alpha}$ are linearly independent
among themselves, the set of HJ equations are related to the characteristic
equations \eqref{eq:12}, whose solutions are trajectories $\xi^{A}=\xi^{A}\left(t^{\alpha}\right)$
in a reduced phase space $\mathbf{T}^{*}\mathbb{Q}_{P}$, parametrized
by $t^{\alpha}$.

The necessary and sufficient conditions for the linear independence
of the parameters $t^{\alpha}$, and therefore for the existence of
the characteristic equations themselves, happen to be the same conditions
for the existence of a complete solution of the HJ equations. In sec.
\ref{sec:3} we introduced the Frobenius' theorem: a system of first-order
PDEs $H'_{\alpha}\left(t^{\beta},\pi_{\beta},q^{a},p_{a}\right)=0$
is completely integrable if, and only if, the functions $H'_{\alpha}$
obey the Frobenius' integrability conditions \eqref{eq:23}. Sufficient
conditions for integrability can, on the other hand, be generalized
to a Poisson algebra \eqref{eq:25}. Since $H'_{\alpha}=0$ are also
canonical constraints, integrability demands these constraints to
be in involution with the PBs. Therefore, a complete set of integrable
HJ equations are also a set of involutive constraints.

If $H'_{\alpha}=0$ is a complete set of involutive constraints, sec.
\ref{sec:4} shows that they are generators of active canonical transformations
in the complete phase space $\mathbf{T}^{*}\mathbb{Q}_{N+1}$. This
is done by building the symplectic structure: first we introduced
the vector fields $X_{\alpha}$, related to the functions $H'_{\alpha}$.
With integrability, the symplectic 2-form $\omega$ splits in two
2-forms $\omega_{P}$ and $a$, eq. \eqref{eq:31}, such that $a$
is identically a null form. In this case, we have shown a known result,
that singular systems have a degenerate symplectic structure, whose
vector fields $X_{\alpha}$ form a basis of the null space of $\omega$,
as can be seen by the relations \eqref{eq:36} and \eqref{eq:37}.
Infinitesimal transformations \eqref{eq:38}, generated by the involutive
constraints, preserve the symplectic structure \eqref{eq:42}, and
therefore Liouville's theorem is implied. Moreover, a canonical action
defined in \eqref{eq:43} is invariant under the so called characteristic
flows \eqref{eq:46}. For constant structure coefficients, the Poisson
algebra \eqref{eq:25} implies the Lie algebra \eqref{eq:29}, and
a Lie group of transformations can be built form the Lie algebra of
the characteristic vector fields.

The connection between the characteristic flows and lagrangian point
(gauge) transformations is discussed in sec. \ref{sub:4.4}. Point
transformations \eqref{eq:47} are CFs in which $\delta t=0$, \emph{i.e.},
time-independent canonical transformations. However, to be canonical,
the transformations themselves must be restricted to $\mathbf{T}^{*}\mathbb{Q}_{P}$,
since their generators $H_{z}^{'}$ must obey their own Poisson algebra
\eqref{eq:48}. Because point transformations are Noether symmetries,
characterized by a set of gauge parameters $\epsilon$, the Lie equation
\eqref{eq:51} can be used to relate $\delta t^{\alpha}$ to this
set. In general, it results in linear dependency between the independent
variables $t^{\alpha}$, as shown directly in the examples of sec.
\ref{sec:5}. We notice that no analogous of Dirac's conjecture is
needed in these results, but the generator of the characteristic flows
depends on a complete set of involutive constraints $H'_{z}$, as
a direct result of Frobenius' theorem. The dependency between the
parameters $t^{\alpha}$, on the other hand, is used to build the
generator of gauge transformations directly from the generator of
the characteristic flows.

\subsection*{Acknowledgments}

M. C. Bertin was partially supported by FAPESP. B. M. Pimentel was
partially supported by CNPq and CAPES. C. E. Valcárcel was supported
by FAPESP.

\end{document}